\documentclass[aps,pra,showpacs,showkeys,floatfix,onecolumn]{revtex4}
\usepackage{epsfig}
\usepackage{graphicx}
\usepackage{amsmath,amssymb,amsfonts,latexsym}
\usepackage{graphicx}

\usepackage{epsfig,amssymb,amsfonts,verbatim}

\usepackage{amsmath}
\usepackage{latexsym}
\usepackage{amsfonts}
\usepackage{amssymb}
\usepackage{color}

\def\bfl{\begin{flushleft}}
\def\efl{\end{flushleft}}
\def\bfr{\begin{flushright}}
\def\efr{\end{flushright}}
\def\bc{\begin{center}}
\def\ec{\end{center}}

\def\ba{\begin{eqnarray}}
\def\ea{\end{eqnarray}}
\def\baa#1{\begin{array}{#1}}
\def\eaa{\end{array}}
\def\bw{\begin{widetext}}
\def\ew{\end{widetext}}
\def\nn{\nonumber }

\def\text#1{\mbox{#1}}

\begin{document}

\title{Many-body Hamiltonian with screening parameter and ionization energy}

\author{Andrew Das Arulsamy}
\email{andrew@physics.usyd.edu.au}


\address{School of Physics, The University of Sydney, Sydney, New South Wales 2006, Australia}


\date{\today}

\begin{abstract}
We prove the existence of a new Hamiltonian that can be used to study strongly correlated matter, which consists of the total energy at temperature equals zero ($E_0$) and the ionization energy ($\xi$) as eigenvalues. We show that the existence of this total energy eigenvalue, $E_0 \pm \xi$, does not violate the Coulombian atomic system. Since there is no equivalent known Hamilton operator that corresponds quantitatively to $\xi$, we employ the screened Coulomb potential operator, which is a function of this ionization energy to analytically calculate the screening parameter ($\sigma$) of a neutral Helium atom in the ground state. In addition, we also show that the energy level splitting due to spin-orbit coupling is inversely proportional to $\xi$ eigenvalue. 
\end{abstract}


\maketitle

\section{Introduction}

Finding even an approximate but an accurate solution to a Coulombian many-body problem (many-electron atoms and solids) is no doubt, one of the most important problems in physics~\cite{stuchi,bogo}. The Rayleigh-Ritz variational (RRV) method~\cite{arfken} is a common and a powerful method regardless of the type of potential used (neglecting $e$-$e$ interaction) to solve a particular Hamiltonian~\cite{hill2,hill3,lyons}. In the presence of $e$-$e$ interaction, one can still use the RRV method, for example, in the presence of screened Coulomb potential (Yukawa-type) to calculate the total energy eigenvalue~\cite{gerry,kobori}. The Yukawa-type potential~\cite{yukawa}, $V = -(1/r)\exp[-\sigma r]$ is one of the well studied potential and the focus is usually on how to enumerate $\sigma$ for a given bound state~\cite{sachs,hulthen,harris,smith,schey,iafrate,rogers,garavelli,gomes,al,castro,naga,du}. In this work however, we will define the screening parameter, $\sigma$ as a function that depends on the ionization potential~\cite{arulsamy}, in which, we will treat $\sigma$ as a many-body parameter in real atoms. On the other hand, there are also methods employed to tackle screening effect namely, renormalization-group~\cite{yukalov}, 1/$N$ expansion~\cite{ashok,amit}, screened Coulomb potential in the momentum representation~\cite{ullah} and, Dirac and Klein-Gordon equations with relativistic effects~\cite{leo,zno,safro}. These methods, though very useful and accurate, do not allow one to associate the different isolated atomic energy levels for all the ions that exist in solids to changes on the electronic excitation probability in solids. This association is important to predict electronic properties of solids with different ions. 

As such, we propose here a new many-body Hamiltonian to study just that, using the Yukawa-type potential. The eigenvalue of this Hamiltonian is made up from the total energy at temperature equals zero ($E_0$) and ionization energy ($\xi$). The total energy from this Hamiltonian has been verified in experiments where the isolated atomic energy levels have been used to predict the changes of the electronic excitation probability in solids (Ref.~\cite{arulsamy} and references therein). Hence, in this paper we do not discuss the application of this Hamiltonian in solids. As such, our motivation here is to give proofs-of-existence that the right-hand-side (eigenvalue) and the left-hand-side (Hamilton operator) of this new many-body Hamiltonian are physically equivalent. We apply this Hamiltonian to Hydrogen and Helium atoms. In order to do so, we will need to find an operator that corresponds to $\xi$ eigenvalue and prove that it yields the same conclusion as $\xi$ itself. We first establish the validity of this eigenvalue for the 1D systems and Hydrogen atom. Secondly, we will show that the screened Coulomb potential ($V_{\rm{sc}}$) is the operator that corresponds to $\xi$, and the screened Coulomb potential in atomic He is also proportional to the ionization energy. We will also derive an expression that associates the spin-orbit coupling and the ionization energy, through the screener, $\sigma$.   

\section{Many-body Hamiltonian as a function of ionization energy}

The 3-dimensional Schrodinger equation for Fermions with mass, $m$ moving in the presence
of potential, $V(\textbf{r})$ is given by (after making use of the linear
momentum operator, $\hat{p} = -i\hbar\nabla^2$)~\cite{beth}.

\begin {eqnarray}
-\frac{\hbar^2}{2m}\nabla^2\varphi = (E + V(\textbf{r}))\varphi. \label{eq:1}
\end {eqnarray}

$E$ denotes the total energy.

We define,
\begin {eqnarray}
\pm\xi := E_{\rm{kin}} - E_0 + V(\textbf{r}), \label{eq:2}
\end {eqnarray}

such that $\pm\xi$ is the energy needed for a particle to overcome
the bound state and the potential that surrounds it. $E_{\rm{kin}}$ and
$E_0$ denote the total energy at $V(\textbf{r})$ = 0 and the energy at $T$
= 0, respectively, i.e., $E_{\rm{kin}}$ = kinetic energy. In physical
terms, $\xi$ is defined as the ionization energy. That is, $\xi$
is the energy needed to excite a particular electron to a finite
$r$, not necessarily $r \rightarrow \infty$.

proof: At temperature, $T$ = 0 and $V(\textbf{r})$ = 0,
$\hat{H}\varphi =
-\frac{\hbar^2}{2m}\nabla^2\varphi =
E_0\varphi$. Hence, from Eq.~(\ref{eq:1}) with $V(\textbf{r})$ = 0 and $T$
= 0, the total energy can be written as

\begin {eqnarray}
E = E_{\rm{kin}} = E_0. \label{eq:4}
\end {eqnarray}

Remark 1: Therefore, for an electron to occupy a higher energy state, $N$
from the initial state, $M$ is more probable than from the initial
state, $L$ if the condition, $[E_N(\geq 0) - E_M(\geq 0)]$ $<$
$[E_N(\geq 0) - E_L(\geq 0)]$ at certain $T$ is satisfied. As for
a hole to occupy a lower state $M$ from the initial state $N$ is
more probable than to occupy a lower state $L$ if the condition,
$|E_M(< 0) - E_N(< 0)|$ $<$ $|E_L(< 0) - E_N(< 0)|$ at certain $T$
is satisfied.

On the other hand, using the above stated new definition
(Eq.~(\ref{eq:2})) and the condition, $T = 0$, we can rewrite the
total energy as

\begin {eqnarray}
E = E_{\rm{kin}} = E_0 \pm \xi.\label{eq:5}
\end {eqnarray}

Therefore, Remark 1 can be rewritten as

Remark 2: For an electron to occupy a higher energy state, $N$ from the
initial state, $M$ is more probable than from the initial state,
$L$ if the condition, $\xi(M \rightarrow N)$ $<$ $\xi(L
\rightarrow N)$ at certain $T$ is satisfied. As for a hole to
occupy a lower state $M$ from the initial state $N$ is more
probable than to occupy a lower state $L$ if the condition, $\xi(N
\rightarrow M)$ $<$ $\xi(N \rightarrow L)$ at certain $T$ is
satisfied.

For electron-like excitations, Remark 2 implies that

\begin {eqnarray}
&&\xi(M \rightarrow N) = [E_N(\geq 0) - E_M(\geq 0)], \nn \\&& \xi(L \rightarrow N) = [E_N(\geq 0) - E_L(\geq 0)], \nn
\end {eqnarray}

while

\begin {eqnarray}
&&\xi(N \rightarrow M) = |E_M(< 0) - E_N(< 0)|, \nn \\&& \xi(N
\rightarrow L)  = |E_L(< 0) - E_N(< 0)|, \nn
\end {eqnarray}

for hole-like excitations. In other words, if we let the energy function, $E = E_0 \pm \xi$ = $E_{\rm{kin}} + V(\textbf{r})$, then we can write the many-electron atomic Hamiltonian as 

\begin {eqnarray}
\hat{H}\varphi = (E_0 \pm \xi)\varphi. \label{eq:7}
\end {eqnarray}

\section{Infinite square-well potential}

Here, the hamiltonian given in Eq.~(\ref{eq:7}) is applied to a
free-electron system, by considering a free-particle of mass $m$
moving in 1-dimension of an infinite square well (width = $a$).
Therefore, Eq.~(\ref{eq:7}) can be solved generally, to give

\begin {eqnarray}
\varphi_n = C \sin\bigg[\bigg(\frac{2m}{\hbar^2}(E_0 \pm
\xi)\bigg)_n^{1/2}x\bigg]. \label{eq:8}
\end {eqnarray}

After normalization, $\int_0^a |C|^2\sin^2[(\frac{2m}{\hbar^2}(E_0
\pm \xi))^{1/2}]dx = |C|^2a/2 = 1$, one obtains $C = \sqrt{2/a}$.
Finally, the normalized wave function,

\begin {eqnarray}
\varphi_n =
\sqrt{\frac{2}{a}}\sin\bigg[\bigg(\frac{2m}{\hbar^2}(E_0 \pm
\xi)\bigg)^{1/2}_nx\bigg]. \label{eq:9}
\end {eqnarray}

Applying the boundary conditions for free-electrons, $V(x)$ = 0;
$\xi = E_{\rm{kin}} - E_0 + [V(x) = 0] = E_{\rm{kin}} - E_0; E = E_0 \pm
\xi$, $\varphi(0)$ = 0 and $\varphi(a)$ = 0 require
$(\frac{2m}{\hbar^2}(E_0 \pm \xi))^{1/2}_n = n\pi/a$. Note here
that the condition, $V(x)$ = 0 that leads to the free-electron
concept also implies the square well potential equals zero
anywhere between 0 and $a$ ($0 < x < a$), and this will stay true
for as long as $a \gg 2r_e$ where $r_e$ denotes an electron's
radius. Eventually, one arrives at

\begin {eqnarray}
\varphi_n = \sqrt{\frac{2}{a}}\sin\bigg[\frac{n\pi}{a}x\bigg].
\label{eq:10}
\end {eqnarray}

Obviously, one can also obtain the exact form of
Eq.~(\ref{eq:10}) from the 1D time-independent Schr\"{o}dinger
equation with $V(x)$ = 0 and $k_n = n\pi/a$ (Ref.~\cite{griffiths5}). Therefore, we can conclude that our Hamiltonian is exactly the same as the usual Hamiltonian. However, in the presence of potential energy, our Hamiltonian will provide additional information on the energetics of Fermions in different atoms. This new information will tell us how one can use the atomic energy level difference to predict the electronic excitation probability in solids via Fermi-Dirac statistics~\cite{arulsamy}. 

\section{1D Dirac-delta potential}

As a matter of fact, there can be many general solutions for
Eq.~(\ref{eq:7}) with $V(x) \neq 0$ and these solutions can be
derived in such a way that they can be compared, term by term with
known wave functions. For example, If $V(x) = -\alpha \delta (x)$,
then we need a solution in the form of $\varphi(x) = C \exp[-iax]$
and the associated wave function can be derived as $\frac{\partial\varphi(x)}{\partial x} = -ia C \exp [-iax],  \frac{\partial^2\varphi(x)}{\partial x^2} =
-a^2\varphi(x)$. we can rewrite Eq.~(\ref{eq:7}) to get

\begin {eqnarray}
&&\frac{\partial^2\varphi}{\partial x^2} = -\frac{2m}{\hbar^2}[E_0
\pm \xi]\varphi, \label{eq:12}
\end {eqnarray}

Using Eq.~(\ref{eq:12}), we find $a = \left[\frac{2m}{\hbar^2}(E_0 \pm \xi)\right]^{1/2}$. Therefore, $\varphi(x) = C \exp\big[-i\big(\frac{2m}{\hbar^2}(E_0 \pm \xi)\big)^{1/2}x\big]$. Normalizing $\varphi$ gives $1 = \int_{-\infty}^{+\infty}|\varphi(x)|^2 dx = \frac{C^2}{i\big(\frac{2m}{\hbar^2}(E_0 \pm \xi)\big)^{1/2}}$. Hence, 

\begin {eqnarray}
&\varphi(x) &= i^{1/2}\bigg[\frac{2m}{\hbar^2}(E_0 \pm
\xi)\bigg]^{1/4}\nn \\&& \times \exp\bigg[-i\bigg(\frac{2m}{\hbar^2}(E_0 \pm
\xi)\bigg)^{1/2}x\bigg].  \label{eq:210}
\end {eqnarray}

Term by term comparison between Eq.~(\ref{eq:210})
and~\cite{griffiths5}

\begin {eqnarray}
\varphi(x) = \frac{m \alpha}{\hbar}\exp\bigg[-\frac{m \alpha
|x|}{\hbar^2}\bigg],\nn
\end {eqnarray}

gives the bound state energy, $E_0 \pm \xi = -m \alpha^2/2\hbar^2$
either by equating $[i^2(\frac{2m}{\hbar^2}(E_0 \pm \xi))]^{1/4}
= m \alpha/\hbar$ or $ -i(\frac{2m}{\hbar^2}(E_0 \pm \xi))^{1/2}x
= -m \alpha |x|/\hbar^2$.

\section{Hydrogen atom}

In this section, Eq.~(\ref{eq:7}) is applied to a Hydrogen atom
with Coulomb potential and subsequently, exact results of its
energy levels are derived. The radial equation of a Hydrogen atom
is given by~\cite{griffiths5}

\begin{eqnarray}
-\frac{\hbar^2}{2m}\frac{d^2u}{dr^2} +
\left[-\frac{e^2}{4\pi\epsilon_0r} +
\frac{\hbar^2}{2m}\frac{l(l+1)}{r^2}\right]u = Eu. \label{eq:440}
\end{eqnarray}

Equation~(\ref{eq:440}) can be rewritten using $E = E_0 \pm
\xi$ and can be readily solved~\cite{griffiths5} to
obtain the principal quantum number, $n$

\begin{eqnarray}
&&2n =
\frac{me^2}{2\pi\epsilon_0\hbar^2\sqrt{\frac{2m}{\hbar^2}(E_0 \pm
\xi)}}. \nn \\&& \therefore \sqrt{\frac{2m}{\hbar^2}(E_0 \pm \xi)}
= \bigg[\frac{me^2}{4\pi\epsilon_0\hbar^2}\bigg]\frac{1}{n} =
\frac{1}{a_Bn}. \label{eq:460}
\end{eqnarray}

$a_B$ denotes the Bohr radius and $\epsilon_0$ is the permittivity
of space. $m$ and $\hbar$ are the particle's mass and the Planck constant, respectively. Equation~(\ref{eq:460}) can be rewritten so as to obtain the energy levels of a Hydrogen atom

\begin{eqnarray}
(E_0 \pm \xi)_n =
-\left[\frac{m}{2\hbar^2}\bigg(\frac{e^2}{4\pi\epsilon_0}\bigg)^2\right]\frac{1}{n^2}
= E_n.\label{eq:470}
\end{eqnarray}

We propose that $(E_0 \pm \xi)_n$ can be rewritten in terms of the standard $E_n$
where, $-(E_0 \pm \xi)_n = - E_n$. Proof:

\begin{eqnarray}
&&-(E_0 \pm \xi)_{m+1} = -\left[E_0 \pm (E_m - E_{m+1})\right]
\Leftrightarrow \rm{Eq.~(\ref{eq:2})} \nn \\&& = -\left[E_m \pm
(E_m - E_{m+1})\right] \Leftrightarrow E_0 = E_{m} = \rm{ground~state}
\nn \\&& = -\left[E_m - E_m + E_{m+1}\right] \Leftrightarrow E_m <
E_{m+1} \therefore \pm \rightarrow - \nn \\&& = -E_m + E_m -
E_{m+1} = -E_{m+1} \label{eq:480}
\end{eqnarray}

Alternatively, one can also show that

\begin{eqnarray}
&&-(E_0 \pm \xi)_{m+1} = -\left[E_0 \pm (E_{m+1} - E_m)\right]
\Leftrightarrow \rm{Eq.~(\ref{eq:2})} \nn \\&& = -\left[E_m \pm
(E_{m+1} - E_m)\right] \Leftrightarrow E_0 = E_{m} = \rm{ground~state}
\nn \\&& = -\left[E_m + E_{m+1} - E_m\right] \Leftrightarrow
E_{m+1} > E_m \therefore \pm \rightarrow + \nn \\&& = -E_m + E_m -
E_{m+1} = -E_{m+1} \label{eq:485}
\end{eqnarray}

Equation~(\ref{eq:480}) or~(\ref{eq:485}) can be used to calculate
$(E_0 \pm \xi)_n$. Example: 

\begin{eqnarray}
&&-(E_0 \pm \xi)_{n=1} = -\left[E_0 \pm (E_{n=1} - E_{n=1})\right]
= -E_{n=1} \nn \\&& -(E_0 \pm \xi)_{n=2} = -\left[E_{n=1} -
(E_{n=1} - E_{n=2})\right] = -E_{n=2} \nn \\&& -(E_0 \pm
\xi)_{n=3} = -\left[E_{n=1} - (E_{n=1} - E_{n=3})\right] =
-E_{n=3} \nn \\&& ... \label{eq:490}
\end{eqnarray}

Consequently, energy levels for a Hydrogen atom can be written
exactly, either as $(E_0 \pm \xi)_n$ or $E_n$. Alternatively, we can also assume a solution for the Hydrogen atom in the form of $\varphi_{n=1,l=0,m=0}(r,b) = C\exp[-br^2]$ comparable with~\cite{griffiths5} the
standard hydrogenic radial wavefunction,

\begin {eqnarray}
\varphi_{100}(r,\theta,\phi) = \frac{1}{\sqrt{\pi a_B^3}}e^{-r/a_B}.
\label{eq:1000}
\end {eqnarray}

Therefore, we obtain $\ln \varphi(r) = \ln C - br \ln e$,
$\frac{1}{\varphi(r)} \frac{\partial \varphi(r)}{\partial r} =
-b$ and $\frac{\partial^2\varphi(r)}{\partial r^2} =
-b\frac{\varphi(r)}{\varphi r}$. Equation~(\ref{eq:12}) is used to obtain 

\begin {eqnarray}
b = \frac{i}{\hbar}\sqrt{2m[E_0 \pm \xi]}.
\label{eq:1001}
\end {eqnarray}

Normalization requires, $1 = \int_{-\infty}^{+\infty}|\varphi(r)|^2 d^3\textbf{r} = 4\pi C^2 \frac{2!}{(2b)^3}$, consequently, $\varphi_{100}(r,b) = \sqrt{\frac{b^3}{\pi}}e^{-br}$ and $b = 1/a_B$. The expectation value of the momentum can be calculated as, $\left\langle p\right\rangle = \int_{-\infty}^{+\infty} \varphi(r,b)(-i\hbar(\partial/\partial r))\varphi(r,b) d^3\textbf{r} = \hbar/ia_B$. The momentum can also be written as, $p = \hbar^2k^2 = \sqrt{2m(E_0 \pm \xi)}$, hence, 

\begin {eqnarray}
&&a_B = \frac{\hbar}{i}\sqrt{\frac{1}{2m(E_0 \pm \xi)}} \nn \\&& \therefore E_0 \pm \xi = -\frac{\hbar^2}{2ma_B^2} = -\frac{m}{2\hbar^2}\bigg(\frac{e^2}{4\pi \epsilon_0}\bigg)^2 = E_1. \label{eq:1004}
\end {eqnarray}

Note that Eq.~(\ref{eq:1004}) can also be obtained from Eq.~(\ref{eq:1001}). Equation~(\ref{eq:1004}) is the ground state energy for atomic Hydrogen, as it should be and it justifies the applicability of Eq.~(\ref{eq:7}). Now, for free many-electron atoms, it is neccessary to invoke Eqs.~(\ref{eq:480}),~(\ref{eq:485}) and~(\ref{eq:490}) which can be easily extended exactly to free
many-electron atoms. For example, the eigenvalue for the atomic Mn with electronic configuration, [Ar]
3d$^5$ 4s$^2$ can be written as (after exciting one of the 4s$^2$ electron to 4p)

\begin{eqnarray}
&&-(E_0 \pm \xi)_{4p^1} = -\left[E_0 \pm (E_{4p^1} -
E_{4s^2})\right] \Leftrightarrow \rm{Eq.~(\ref{eq:2})} \nn \\&& =
-\left[E_{4s^2} \pm (E_{4p^1} - E_{4s^2})\right] \Leftrightarrow
E_0 = E_{4s^2} = \rm{initial~state} \nn \\&& = -\left[E_{4s^2} +
E_{4p^1} - E_{4s^2}\right] \Leftrightarrow E_{4p^1}
> E_{4s^2} \therefore \pm \rightarrow + \nn \\&& = -E_{4p^1} \label{eq:500}
\end{eqnarray}

Using Eq.~(\ref{eq:2}), one can also write, $\xi = E_{\rm{kin}} -
E_0 + V(x) = E_{4p^1} - E_{4s^2}$, which suggests that we can
describe the excited electron's properties from the energy level
difference, before and after the excitation. All the results presented
here, thus far are exact and straight forward.

\section{Screened Coulomb potential}

In the previous sections, we only worked on the right-hand-side (RHS) of the Hamiltonian, Eq.~(\ref{eq:7}), which is the total energy eigenvalue and we did not touch the Hamilton operator (LHS). In the subsequent sections however, we will need to solve the Hamilton operator in order to evaluate the influence of the ionization energy on the operator side (LHS), whether they tell us the same story. Of course, we cannot expect to find an isolated or a stand-alone operator that could correspond quantitatively to the ionization energy eigenvalue on the RHS. This is true for all atoms with potential energy attached, including the electron-electron ($e$-$e$) interaction term that gives rise to screening. Mathematically, we say that this one(LHS)-to-one(RHS) ($V_{\rm{sc}}(\textbf{r})$ to $\xi$) correspondence does not exist because $\xi$ has been defined as a function of both kinetic and potential energies (see Eq.~(\ref{eq:2})). However, we can repackage the Coulombian $e$-$e$ repulsion potentials as the screened Coulomb potential (Yukawa-type) for atoms other than Hydrogen. Therefore, we can use the screened Coulomb potential to evaluate how the ionization energy influences the screened Coulomb potential in atomic He. Before we move on, let us just accentuate an important point here. Indeed, it is true that common techniques of many-body theory require handling of the Hamilton operator first, and then calculate the corresponding eigenvalue, ($E$, the total energy). In this paper however, we are doing just the opposite, meaning, we have redefined the total energy eigenvalue, as given in Eq.~(\ref{eq:7}), and now we are in the midst of verifying its consequences on the Hamilton operator. Of course, this verification is indirect simply because we do not have a stand-alone operator that quantitatively corresponds to the eigenvalue, $\xi$. As such, we will make use of the screened Coulomb potential operator, which is given by~\cite{arulsamy} 

\begin {eqnarray}
\hat{V}_{\rm{sc}} = \frac{e}{4\pi\epsilon_0r}e^{-\mu re^{\frac{1}{2}\lambda(- \xi)}} = \frac{e}{4\pi\epsilon_0r}e^{-\sigma r}. \label{eq:11c}
\end {eqnarray}

Here, $\mu$ is the screener's constant of proportionality, while $\lambda = (12\pi\epsilon_0/e^2)a_B$. All we have to do now is to show that for atomic He, $\hat{\left\langle V\right\rangle}_{\rm{sc}}$ is inversely proportional to $\sigma$. Secondly, we will also need to show that $\hat{\left\langle V\right\rangle}_{\rm{sc}}$ is proportional to the ionization energy, by comparing with other two-electron system. For example, when we compare atomic He with other two-electron ions, namely, Li$^+$ and Be$^{2+}$, then we expect, $\hat{\left\langle V\right\rangle}_{\rm{sc}}^{\rm{He}}$ $<$ $\hat{\left\langle V\right\rangle}_{\rm{sc}}^{\rm{Li^+}}$ $<$ $\hat{\left\langle V\right\rangle}_{\rm{sc}}^{\rm{Be^{2+}}}$, because $\xi_{\rm{He}}(54.4-24.6 = 29.8$ eV) $<$ $\xi_{\rm{Li^+}}(122.5-73.6 = 46.8$ eV) $<$ $\xi_{\rm{Be^{2+}}}(217.7-153.9 = 63.8$ eV). 

\section{Helium atom}

\subsection{Many-body Hamiltonian and screened Coulomb potential}

Here, we will first find the expectation value for the screened Coulomb potential for atomic He. The He wavefunction is the product of two hydrogenic wavefunctions (recall that we are not solving the Hamiltonian for the He atom, rather we are only solving the screened Coulomb potential operator)~\cite{beth}, 

\begin {eqnarray}
\varphi_0(\textbf{r}_1,\textbf{r}_2) = \varphi_{100}(\textbf{r}_1)\varphi_{100}(\textbf{r}_2) = \frac{Z^3}{a_B^3\pi}e^{-Z(r_1 + r_2)/a_B}.\label{eq:13c}
\end {eqnarray}
 
Subsequently, the mutual screening between the electrons, $\textbf{r}_1$ and $\textbf{r}_2$ implies that both electrons have identical effective charge of $<$ 2$e$ because they screen each other. Hence,    

\begin {eqnarray}
&&\hat{\left\langle V\right\rangle}_{\rm{sc}} \nn \\&&= \frac{e^2}{4\pi \epsilon_0}\bigg[\frac{Z^3}{a_B^3\pi}\bigg]^2 \int \frac{1}{|\textbf{r}_1-\textbf{r}_2|}e^{-2Z(r_1 + r_2)/a_B} e^{-\sigma (r_1 + r_2)} d^3\textbf{r}_1 ~d^3\textbf{r}_2. \nn \\&& = \frac{e^2}{4\pi\epsilon_0}\bigg(\frac{1}{2a_B}\bigg)\frac{40Z^6}{(2Z+a_B\sigma)^5} = \frac{40Z^6E_1}{\big[2Z+a_B\sigma]^5}, \label{eq:14c}
\end {eqnarray}

where, $|\textbf{r}_1 - \textbf{r}_2| = \sqrt{r_1^2 + r_2^2 - 2r_1r_2\cos(\theta_2)}$. From Eq.~(\ref{eq:14c}), it is clear that $\sigma$ is inversely proportional to the ionization energy, whereas the screened Coulomb potential is proportional to the ionization energy. The complete Hamiltonian for the Helium atom can be written as (with the screened Coulomb potential as the correction term, ignoring fine structure and other corrections)

\begin {eqnarray}
\hat{H} = \hat{H}_{\rm{o}} + \hat{V}_{\rm{sc}} .\label{eq:15cc}
\end {eqnarray}

Where $\hat{H}_{\rm{o}}$ is given by~\cite{beth,griffiths5}

\begin {eqnarray}
\hat{H}_{\rm{o}} = -\frac{\hbar^2}{2m}(\nabla_1^2 + \nabla_1^2) -\frac{e^2}{4\pi \epsilon_0}\bigg[ \frac{2}{r_1} + \frac{2}{r_2} \bigg ] .\label{eq:15c}
\end {eqnarray}

In Eq.~(\ref{eq:15cc}), we again simplified the Hamilton operator by strictly forcing the $e$-$e$ interaction into the $\hat{\left\langle V\right\rangle}_{\rm{sc}}$ as the correction term. Therefore, the non-relativistic many-body Hamiltonian (again, ignoring the fine structure corrections) can now be written as

\begin {eqnarray}
\hat{H} = -\frac{\hbar^2}{2m}\sum_i \nabla_i^2 -\frac{e^2}{4\pi \epsilon_0}\frac{1}{2}\sum_{i\neq j}\bigg[\frac{Z}{r_i} - \frac{1}{|\textbf{r}_i-\textbf{r}_j|}e^{-\sigma (r_i + r_j)}\bigg],\label{eq:1abc}
\end {eqnarray} 

while its eigenvalue is exactly, equals to $E_0 \pm \xi$. Firstly, the aim of Eq.~(\ref{eq:1abc}) is to point out that the eigenvalue, $\xi$ and $\hat{\left\langle V\right\rangle}_{\rm{sc}} $ provide the same conclusions. Secondly, one can use the ionization energy to make certain quantitative and qualitative predictions in solids~\cite{arulsamy}, as well as in atoms (discussed here and in the subsequent sections). The total-energy expectation value for atomic He can be obtained (as given below) by using Eq.~(\ref{eq:1abc}), 

\begin {eqnarray}
\hat{\left\langle H\right\rangle} = 8E_1 + \hat{\left\langle V\right\rangle}_{\rm{sc}}. \label{eq:16c}
\end {eqnarray}
  
Note here that Eq.~(\ref{eq:16c}) is mathematically identical with Eq.~(\ref{eq:7}), where $E_0 \pm \xi = 8E_1 + \hat{\left\langle V\right\rangle}_{\rm{sc}}$ and $\xi$ physically corresponds to $\hat{\left\langle V\right\rangle}_{\rm{sc}}$ through the screener, $\sigma$ as defined earlier. Our problem now is to solve Eq.~(\ref{eq:16c}), which is given by 

\begin {eqnarray}
\hat{\left\langle H\right\rangle} = 8E_1 + \frac{40Z^6E_1}{\big[2Z+a_B\sigma_{\rm{He}}]^5}. \label{eq:17c}
\end {eqnarray}

Equations~(\ref{eq:17c}) tells us that with increased ionization energy ($\xi$), the magnitude of the screener ($\sigma$) is reduced and eventually, gives rise to stronger $e$-$e$ interaction, $\hat{\left\langle V\right\rangle}_{\rm{sc}}$. This conclusion is understandable because small $\xi$ implies weak Coulomb force between the valence electron and the atomic core electrons, which in turn gives a strongly screened valence electron. As a consequence, this particular valence electron interacts less with the core electrons and defines the limit between the strongly correlated electrons ($\xi \rightarrow \infty$) and free-electron solids ($\xi \rightarrow 0$). Using the RRV method, one can capture the exact scenario stated above with variationally determined charge strength, $Ze < 2e$ for He atom~\cite{beth,griffiths5}. $Z$ is the atomic number. Since the ground state energy of the neutral He and H are known~\cite{ken,koga}, 79.005 eV and 13.600 eV, thus we can estimate the screener for atomic He,

\begin {eqnarray}
&\sigma_{\rm{He}} &= \frac{1}{a_B}\bigg[\bigg(\frac{40(2^6)E_1}{-79.005 + 8E_1}\bigg)^{\frac{1}{5}}-4\bigg] \nn \\&& = 2.02 \times 10^{9} ~\rm{m}^{-1}. \label{eq:17e}
\end {eqnarray}

The limit, $\lim_{\xi \rightarrow \infty}\hat{\left\langle V\right\rangle}_{\rm{sc}}$ = $\lim_{\sigma \rightarrow 0}$ $\hat{\left\langle V\right\rangle}_{\rm{sc}}$ = 34 eV, purely due to electron-electron interaction (stronger) with zero screening. This value agree exactly with the one calculated in Ref.~\cite{griffiths5}, also without screening. As a consequence, the ionization energy theory does not violate the true physical picture of the atomic He, both quantitatively and qualitatively. Now, let us calculate the screened Coulomb potential for other two-electron ions, namely, Li$^+$ and Be$^{2+}$. Using Eq.~(\ref{eq:14c}), in the limit $\sigma \rightarrow 0$, we obtain, $\hat{\left\langle V\right\rangle}_{\rm{sc}}^{\rm{Li^+}} = 51$ eV and $\hat{\left\langle V\right\rangle}_{\rm{sc}}^{\rm{Be^{2+}}} = 68$ eV. Hence, we have $\hat{\left\langle V\right\rangle}_{\rm{sc}}^{\rm{He}} (34$ eV) $<$ $\hat{\left\langle V\right\rangle}_{\rm{sc}}^{\rm{Li^+}} (51$ eV) $<$ $\hat{\left\langle V\right\rangle}_{\rm{sc}}^{\rm{Be^{2+}}} (68$ eV), as expected. Furthermore, we can countercheck whether Eq.~(\ref{eq:17c}) is derived without mistakes by showing that it satisfies the Hellmann-Feynman theorem~\cite{hell,feyn},

\begin {eqnarray}
&&\frac{\partial \hat{\left\langle H\right\rangle}}{\partial \sigma} = \left\langle \varphi_0\left|\frac{\partial \hat{H}}{\partial \sigma}\right| \varphi_0\right\rangle \nn \\&& 40Z^6E_1\frac{\partial}{\partial \sigma}\bigg[\frac{1}{(2Z+a_B\sigma)^5}\bigg] = \frac{e^2}{4\pi \epsilon_0}\left\langle \varphi_0\left|\frac{\partial }{\partial \sigma}\bigg(\frac{e^{-\sigma (r_1 + r_2)}}{|\textbf{r}_1 - \textbf{r}_2|}\bigg)\right| \varphi_0\right\rangle \nn \\&& = -\frac{200Z^6E_1 a_B}{(2Z+a_B\sigma)^6}, \label{eq:18c}
\end {eqnarray}

In summary, we have justified the influence of the ionization energy in Eq.~(\ref{eq:7}) through the screeners and the screened Coulomb potential, applied to the Helium atom via Eq.~(\ref{eq:17c}). Equation~(\ref{eq:7}) is technically easier to use, for example, since $\xi$ is unique for each atom and using the total energy, $E_0 \pm \xi$ as one of the restrictive condition, we can derive the ionization energy based Fermi-Dirac statistics, which can be applied to non free-electronic solids. The respective distributions for electron and hole are given by~\cite{arulsamy}

\begin{eqnarray}
&&f_e(E_0,\xi) = \frac{1}{e^{[\left(E_0 + \xi
\right) - E_F^{(0)}]/k_BT }+1}, \nn
\\&& f_h(E_0,\xi) = \frac{1}{e^{[E_F^{(0)} - \left(E_0 - \xi
\right)]/k_BT}+1}. \label{eq:360}
\end{eqnarray}

where, for solids, $E_F^{(0)}$ is the Fermi level at $T$ = 0, $T$ and $k_B$ are the temperature and Boltzmann constant, respectively. 

\subsection{Energy level splitting due to spin-orbit coupling}

In this section, our intention is to find a qualitative expression between the spin-orbit splitting and the ionization energy. The spin-orbit coupling operator after incorporating the screened Coulomb potential (Eq.~(\ref{eq:11c})) is given by   

\begin {eqnarray}
&\hat{H}_{\rm{SOC}} &= \frac{\hat{\textbf{S}}\cdot\hat{\textbf{L}}}{2m^2c^2}\frac{Ze^2}{4\pi\epsilon_0}\left\{\frac{1}{r}\frac{d}{dr}\frac{1}{r}e^{-\mu re^{\frac{1}{2}\lambda(-\xi)}}\right\} \nn \\&& = \hat{\textbf{D}}\bigg\{\frac{\sigma e^{-\sigma r}}{r^2}+\frac{e^{-\sigma r}}{r^3}\bigg\}, \label{eq:20d}
\end {eqnarray}

where~\cite{schwabl}

\begin {eqnarray}
&\hat{\left\langle \textbf{D}\right\rangle}& = \frac{Ze^2\hbar^2}{16m^2c^2\pi\epsilon_0} (l, -l-1), \label{eq:20e}
\end {eqnarray}

$c$ is the speed of light, $\hat{\textbf{S}}$ and $\hat{\textbf{L}}$ are the spin and orbital angular momentum operators, respectively. The $l$ in $(l, -l-1)$ is for $j$ = $l + \frac{1}{2}$, while the $-l-1$ is for $j$ = $l - \frac{1}{2}$. We ignore the irrelevant part of the Hamiltonian for the time being (labeled $\hat{\textbf{D}}$) and focus on the term in the curly bracket. Here we consider the Hydrogenic wavefunction again, given by~\cite{beth}

\begin {eqnarray}
&&\varphi_{210} (\textbf{r}_1,\textbf{r}_2) \nn \\&& = \frac{1}{32\pi}\bigg(\frac{Z}{a_B}\bigg)^5r_1r_2e^{-\frac{Z}{a_B}(r_1+r_2)}\cos\theta_1\cos\theta_2. \label{eq:21d}
\end {eqnarray}

Recall here that we can use the appropriate Hydrogenic wavefunction, depending on the type of orbital, since our corrections have been incorporated into the Screened Coulomb potential (see Eq.~(\ref{eq:16c})). From Eq.~(\ref{eq:20d}) we have,

\begin {eqnarray}
\hat{\left\langle H\right\rangle}_{\rm{SOC}} &=& \hat{\left\langle \textbf{D}\right\rangle}\bigg[\left\langle \varphi_{210} \left|\frac{\sigma e^{-\sigma r}}{r^2}\right|\varphi_{210}\right\rangle \nn \\&& + \left\langle \varphi_{210} \left| \frac{e^{-\sigma r}}{r^3} \right| \varphi_{210} \right\rangle\bigg]. \label{eq:22d}
\end {eqnarray}

We solve the two potential terms separately,
  
\begin {eqnarray}
&&\hat{\left\langle \textbf{D}\right\rangle}\cdot \left\langle \varphi_{210} \left|\frac{\sigma e^{-\sigma r}}{r^2}\right|\varphi_{210}\right\rangle \nn \\&& = \hat{\left\langle \textbf{D}\right\rangle}\frac{\sigma}{(32\pi)^2}\bigg(\frac{Z}{a_B}\bigg)^{10}\int r_1^2e^{-\frac{Z}{a_B}r_1}\cos^2\theta_1~d^3\textbf{r}_1 \nn \\&& \times \int r_2^2\frac{e^{-[\frac{Z}{a_B}+\sigma]r_2}}{r_2^2}~\cos^2\theta_2~d^3\textbf{r}_2 \nn \\&& = \hat{\left\langle \textbf{D}\right\rangle}\bigg(\frac{Z}{a_B}\bigg)^5\frac{\sigma a_B^3}{12(\sigma a_B + Z)^3}, \label{eq:23d}
\end {eqnarray}

and 

\begin {eqnarray}
\hat{\left\langle \textbf{D}\right\rangle}\cdot \left\langle \varphi_{210} \left| \frac{e^{-\sigma r}}{r^3} \right| \varphi_{210} \right\rangle = \hat{\left\langle \textbf{D}\right\rangle}\bigg(\frac{Z}{a_B}\bigg)^5\frac{a_B^2}{24(\sigma a_B + Z)^2}.\nn  
\end {eqnarray}

Finally,

\begin {eqnarray}
\hat{\left\langle H\right\rangle}_{\rm{SOC}} = \hat{\left\langle \textbf{D}\right\rangle}\bigg(\frac{Z}{a_B}\bigg)^5\bigg[\frac{\sigma a_B^3}{12(\sigma a_B + Z)^3} + \frac{a_B^2}{24(\sigma a_B + Z)^2}\bigg]. \nn \\&& \label{eq:25d}
\end {eqnarray}

Apart from $\hat{\left\langle V\right\rangle}_{\rm{sc}}$, we also find that the spin-orbit separation, $\hat{H}_{\rm{SOC}}$ is proportional to $\xi$. The $\hat{H}_{\rm{SOC}}$ also satisfies the Hellmann-Feynman theorem,

\begin {eqnarray}
\frac{\partial \hat{\left\langle H\right\rangle}_{\rm{SOC}}}{\partial \sigma} &&= \left\langle \varphi_{210}\left|\frac{\partial \hat{H}_{\rm{SOC}}}{\partial \sigma}\right| \varphi_{210}\right\rangle \nn \\&& = -\hat{\left\langle \textbf{D}\right\rangle}\frac{\sigma a_B^4}{4(\sigma a_B + Z)^4}\bigg(\frac{Z}{a_B}\bigg)^5. \label{eq:18g}
\end {eqnarray}

Using the value for screener calculated earlier (2.0 $\times 10^{9}$ m$^{-1}$), we find a crude estimate for the spin-orbit separation for He in 2p-orbital, which is given by,

\begin {eqnarray}
&\delta\hat{\left\langle H\right\rangle}_{\rm{SOC}} & = 3.9 \times 10^{-6} (l, -l-1)~\rm{eV} \nn \\&& = 0.72~\rm{meV} ~\Leftrightarrow ~ \textit{l} = 1. \label{eq:26d} 
\end {eqnarray}

This calculated value is reasonable compared to the relativistic corrections~\cite{ken} for 1s orbital ($l$ = 0), which is 0.00311 eV. Apparently, $\delta\hat{\left\langle H\right\rangle}_{\rm{SOC}}$ will be smaller even for $l$ = 1 as a result of relativistic kinetic energy exclusion, because relativistic correction for $l$ = 1 must come from both spin-orbit coupling and relativistic kinetic energy~\cite{beth}.     

\section{Application to many-electron atoms}

There is one important relationship, given in Eq.~(\ref{eq:25d}) that we can use it to qualitatively check the energy level splitting in many-electron atoms. Equation~(\ref{eq:25d}) says that the magnitude of the energy level splitting ($\delta$), of a particular orbital is proportional to the energy level difference ($\xi$), of two different orbitals. That is, if the energy level difference is large, then it implies that $\delta$ is also large. The $\delta$ and $\xi$ can be compared to the experimental atomic spectra using the following approximate formulae, 

\begin {eqnarray}
\xi =~ ^{\rm{y}}E_L^{\rm{max}} -~ ^{\rm{x}}E_L^{\rm{min}}, \label{eq:27d} 
\end {eqnarray}
and         
\begin {eqnarray}
\delta =~ ^{\rm{x}}E_L^{\rm{max}} -~ ^{\rm{x}}E_L^{\rm{min}}, \label{eq:28d} 
\end {eqnarray}

where x and y denote the different electronic configuration (EC) and/or spectroscopic term. That is, for atomic He, x = 1s3p:$^3$P and y = 1s3d:$^3$D, for atomic C, x = 2s$^2$2p$^2$:$^3$P and y = 2s$^2$2p$^2$:$^1$D, and so on. The magnitude of $\xi$ and $\delta$ can be calculated for atomic Mn using Eqs.~(\ref{eq:27d}) and~(\ref{eq:28d}): $^{\rm{y}}E_L^{\rm{max}}$ =  2.319170 eV, $^{\rm{x}}E_L^{\rm{min}}$ = 2.114214 eV, therefore $\xi$ = 2.319170 $-$ 2.114214 = 204.96 meV. Whereas, $^{\rm{x}}E_L^{\rm{max}}$ = 2.186728 eV thus, $\delta$ = 2.186728 $-$ 2.114214 = 72.51 meV. Here, three atomic spectra namely, He-I, C-I and Mn-I, as representatives of the periodic table are considered and are available in Ref.~\cite{nist}. However, there are also atoms where the energy level difference is not clearly defined because of the energy-level overlapping, and incomplete spectroscopic data (for example, atomic Nd-I). The selected electronic configuration (EC), $J$ (total angular momentum), spectroscopic term, $\xi$, energy levels ($E_L$) and its splittings ($\delta$) are listed in Table~\ref{Table:II}. It is clear from Table~\ref{Table:II} that the ionization energy concept invoked in the many-body Hamiltonian (see Eqs.~(\ref{eq:7}) and~(\ref{eq:1abc})) explains the energy level splitting in many-electron atoms where, the energy level splitting, $\delta$ is proportional to the energy level difference of a particular atom.       


\begin{table}[ht]
\caption{The energy level difference ($\xi$) and the corresponding energy level splitting ($\delta$) were calculated using the experimental data obtained from Ref.~\cite{nist}, Eqs.~(\ref{eq:27d}) and~(\ref{eq:28d}).} 
\begin{tabular}{l c c c c c} 
\hline\hline 
\multicolumn{1}{l}{EC}              & Term             &   $J$        & $E_L$ (eV)    & $\xi$ (meV) & $\delta$ (meV) \\  
\hline 
He: 1s3p                            &  $^3$P$^{\rm{o}}$&  2           &  23.0070718   & 66.58 & 0.036 \\ 
                                    &                  &  1           &  23.0070745   &       &       \\
                                    &                  &  0           &  23.0071081   &       &       \\

He: 1s3d                            &  $^3$D           &  1           &  23.0736551   &       &       \\ \hline 

He: 1s4p                            &  $^3$P$^{\rm{o}}$&  2           &  23.7078898   & 28.20 & 0.015 \\ 
                                    &                  &  1           &  23.7078909   &       &       \\ 
                                    &                  &  0           &  23.7079046   &       &       \\ 

He: 1s4d                            &  $^3$D           &  1           &  23.7360912   &       &       \\ \hline

C: 2s$^2$2p$^2$                     &  $^3$P           &  0           &  0.000000     &1263.73& 5.380 \\ 
                                    &                  &  1           &  0.002033     &       &       \\ 
                                    &                  &  2           &  0.005381     &       &       \\ 

C: 2s$^2$2p$^2$                     &  $^1$D           &  2           &  1.263725     &       &       \\  \hline

C: 2s2p$^3$                         &  $^3$D$^{\rm{o}}$&  3           &  7.945765     & 591.33& 0.500 \\ 
                                    &                  &  1           &  7.946128     &       &       \\ 
                                    &                  &  2           &  7.946265     &       &       \\ 

C: 2s$^2$2p($^2$P$^{\rm{o}}$)3p     &  $^1$P           &  1           &  8.537096     &       &       \\ \hline                    

Mn: 3d$^6$($^5$D)4s                 &  a$^6$D                 & $\frac{9}{2}$ &  2.114214     & 204.96& 72.51 \\ 
                                    &                         & $\frac{7}{2}$ &  2.142695     &       &       \\                     
                                    &                         & $\frac{5}{2}$ &  2.163713     &       &       \\ 
                                    &                         & $\frac{3}{2}$ &  2.178214     &       &       \\                     
                                    &                         & $\frac{1}{2}$ &  2.186728     &       &       \\ 

Mn: 3d$^5$($^6$D)4s4p($^3$P$^{\rm{o}}$)  & z$^8$P$^{\rm{o}}$  & $\frac{9}{2}$ &  2.319170     &       &       \\\hline

Mn: 3d$^6$($^5$D)4s                      & a$^4$D             & $\frac{7}{2}$ &  2.888419     & 186.67& 64.74 \\
                                         &                    & $\frac{5}{2}$ &  2.919729     &       &       \\
                                         &                    & $\frac{3}{2}$ &  2.940845     &       &       \\
                                         &                    & $\frac{1}{2}$ &  2.953163     &       &       \\ 

Mn: 3d$^5$($^6$D)4s4p($^3$P$^{\rm{o}}$)  & z$^6$P$^{\rm{o}}$  & $\frac{7}{2}$ &  3.075087     &       &       \\        \hline  
\end{tabular}
\label{Table:II} 
\end{table}

\section{Conclusions}

In conclusion, we have established a new many-body Hamiltonian where the screened Coulomb potential operator is found to correspond directly qualitatively with the ionization energy eigenvalue. It is not possible to find an operator that corresponds quantitatively due to the definition of the ionization energy itself. The advantage of working with this Hamiltonian is that we do not need to solve the Hamiltonian in order to make qualitative predictions in many-electron atoms. In other words, we can use the total energy, which is the function of ionization energy in order to predict the changes in screening strength and spin-orbit separation for different atoms. Here, we have given proofs of existence by applying it to one-dimensional systems, atomic hydrogen and helium, and ionic lithium and beryllium. We found that the ionization energy is a mathematically and physically valid eigenvalue and its corresponding screened Coulomb potential operator is also found to be valid. 

\section*{Acknowledgments}

I would like to thank the School of Physics, University of Sydney and Professor Catherine Stampfl for the USIRS award. Special thanks to Ronie Entili for finding Ref.~\cite{hell} and Simone Piccinin for editing the introduction.

\end{document}